\documentclass[aps, prl, 10pt, superscriptaddress, preprintnumbers,twocolumn]{revtex4-1}

\usepackage{amsmath,amssymb,mathrsfs}
\usepackage{graphicx}
\usepackage[mathscr]{euscript}

\newcommand{\la}{\label}
\newcommand{\bbm}{\begin{multline}}
\newcommand{\eem}{\end{multline}}
\newcommand{\be}{\begin{equation}}
\newcommand{\ee}{\end{equation}}
\newcommand{\bea}{\begin{eqnarray}}
\newcommand{\eea}{\end{eqnarray}}
\newcommand{\p}{\partial}

\newcommand{\comment}[1]{}

\def\lc{\left\lfloor}   
\def\rc{\right\rfloor}

\usepackage{color}

\allowdisplaybreaks[1]

\begin{document}

\title{Boundary effective action for quantum Hall states}

\preprint{YITP-SB-15-18}

\author{Andrey~Gromov}
\affiliation{Department of Physics and Astronomy, Stony Brook University,  Stony Brook, NY 11794, USA}

\author{Kristan~Jensen}
\affiliation{C.N.~Yang Institute for Theoretical Physics, SUNY Stony Brook,  Stony Brook, NY 11794, USA}

\author{Alexander G.~Abanov}
\affiliation{Department of Physics and Astronomy, Stony Brook University,  Stony Brook, NY 11794, USA}
\affiliation{Simons Center for Geometry and Physics, Stony Brook University,  Stony Brook, NY 11794, USA}

\date{\today}

\begin{abstract}
We consider quantum Hall states on a space with boundary, focusing on the aspects of the edge physics which are completely determined by the symmetries of the problem. There are four distinct terms of Chern-Simons type that appear in the low-energy effective action of the state. Two of these protect gapless edge modes. They describe Hall conductance and, with some provisions, thermal Hall conductance. The remaining two, including the Wen-Zee term, which contributes to the Hall viscosity, do not protect gapless edge modes but are instead related to local boundary response fixed by symmetries. We highlight some basic features of this response. It follows that the coefficient of the Wen-Zee term can change across an interface without closing a gap or breaking a symmetry.
\end{abstract}


\maketitle

\paragraph{Introduction.} 

Topology and geometry play an important role in modern condensed matter physics. For example, in quantum Hall systems, the observed quantization and rigidity of the Hall conductance $\sigma_H$ are most naturally explained using topological arguments~\cite{TKNN}. There are several types of topology at play in this example. In particular, the Hall conductance appears as the coefficient in front of a Chern-Simons (CS) term in the bulk low-energy effective action $S_{bulk}$ of the state,
\be
 \label{E:AdA}
	S_{bulk} = \frac{\sigma_H}{2}\int_{\mathcal{M}} d^3x\,\,
	\epsilon^{\mu\nu\rho} A_{\mu} \partial_{\nu}A_{\rho} + \hdots\,,
\ee
where $A_{\mu}$ is an external electromagnetic gauge field and $\mathcal{M}$ the three-dimensional space-time. Charge conservation then implies that $\sigma_H$ cannot vary continuously in space or time, and is quantized in a way that depends on the electric charges of quasiparticles. 

This CS term has another property: it is gauge-invariant up to a boundary term, and so is invariant on a closed spacetime, but not on a spacetime with a boundary. This non-invariance cannot be cured by adding local boundary terms built from $A_\mu$ and its derivatives. Charge conservation together with the existence of the CS term then imply that there is a gapless, non-gauge-invariant edge theory which cancels the non-invariance of the bulk. Namely, the quantum effective edge action $S_{edge}$ living on the spacetime boundary $\p \mathcal{M}$ obeys
\be
 \label{E:inflow}
	\delta_{\Lambda} S_{edge} = - \delta_{\Lambda} S_{bulk} 
	= - \frac{\sigma_H}{2}\int_{\partial\mathcal{M}}d^2x\, \Lambda\,\epsilon^{\alpha\beta}\partial_{\alpha} A_{\beta}\,,
\ee
where $\Lambda$ is the gauge transformation parameter and $\alpha,\beta$ are boundary indices. This non-invariance of the edge theory is known as an ``anomaly,'' and its cancellation against the variation of a CS term is an example of ``anomaly inflow''~\cite{callan1985anomalies}.
The edge depends on the details of the state, including boundary conditions, and is often unknown, but it must possess the anomaly~\eqref{E:inflow} and be gapless so as to make up for the non-invariance of the bulk at arbitrarily low energies.

There are other rigid transport coefficients in quantum Hall states. These are encoded in the dimensionless coefficients of CS terms in the low-energy action of the state~\cite{WenZeeShiftPaper,1993-frohlich,2000-ReadGreen, 2009-Read-HallViscosity,2009-Haldane-HallViscosity, 2011-ReadRezayi, bradlyn-read-2012kubo, Abanov-2014}. The most well-known of these is the Hall viscosity \cite{1995-AvronSeilerZograf} and it is related to the Wen-Zee (WZ) term~\cite{WenZeeShiftPaper, 2012-HoyosSon}, which we discuss below. This term is not invariant on a spacetime with boundary. One natural question is: does the WZ term protect the existence of gapless edge modes, or instead correspond to some boundary-localized response?

The goal of this Letter is to answer this question. We consider CS terms consistent with the symmetries of a quantum Hall state, and deduce which correspond to anomalies and which to local boundary terms. We show that Wen-Zee terms belong to the latter category and do not correspond to protected gapless edge states. Nevertheless, they still encode symmetry-protected boundary response, which we discuss below. Our analysis only employs the symmetries of the problem as in e.g.~\cite{2012-HoyosSon, Gromov-galilean, gromov-thermal, bradlyn2014low}, and so is robust even when the microscopic system underlying the Hall state is strongly interacting.
   
\paragraph{The setup.} 

We consider gapped systems in two spatial dimensions with a conserved current $j^{\mu}$ and spatial stress tensor $T^{ij}$, to which we respectively couple an external gauge field $A_{\mu}$ and spatial metric $g_{ij}$. We assume that the underlying state is rotationally invariant in flat space~\footnote{The precise statement is that we consider systems which depend on only one spatial metric, as compared with~\cite{2009-Haldane-HallViscosity}. The generalization of this work to systems with more than one spatial metric is immediate.}. Due to the gap, the low-energy effective action $S_{bulk}$ only depends on the external fields $(A_{\mu},g_{ij})$ and can be presented as an expansion in gradients thereof.

The total low-energy effective action $S_{eff} = S_{edge} + S_{bulk}$ is invariant under all the symmetries of the underlying theory, including gauge transformations under which $A_{\mu}$ varies as $\delta_{\Lambda}A_{\mu}=\partial_{\mu}\Lambda$. It is also invariant under spatial reparameterizations of space $x^i=x^i(y^j)$, provided that we equip the external fields $(A_{\mu},g_{ij})$ with the right transformation properties. We will use these symmetries to constrain the form of both bulk and boundary parts of the effective action.

One can extend the spatial reparameterization invariance to a full space-time invariance by introducing a frame $\beta^{\mu}_a=(\beta^{\mu}_0,E^{\mu}_A)$ and coframe $(\beta^{-1})^b_{\nu}=((\beta^{-1})^0_{\nu},e_{\nu}^B)$, which we have separated into temporal and spatial parts. Here $\mu,\nu$ are spacetime indices, $a,b=0,1,2$ order the basis, and $A,B=1,2$ label spatial vectors. (A frame is just a local basis of tangent vectors.)  We take the ``time vector'' to be $\beta^{\mu}_0 = \delta_t^{\mu}$ and $(\beta^{-1})_{\mu}^0=\delta_{\mu}^t$. The remaining spatial vectors $E_A$ with $A=1,2$ give a spatial vielbein and the $e^A_{\mu}$ a spatial coframe. From the $e^A_{\mu}$ we construct a spacetime covariant version of $g_{ij}$, given by $g_{\mu\nu}=\delta_{AB}e^A_{\mu}e^B_{\nu}$, which is invariant under local $SO(2)$ rotations which rotate the $e^A_{\mu}$ into each other. We use an $SO(2)$ spin connection for this transformation, $\omega_{\mu}=\frac{1}{2}\epsilon^A{}_B E_A^{\nu}D_{\mu} e_{\nu}^B$, which characterizes the geometry. Here $D_{\mu}$ is a covariant derivative defined with a connection $\Gamma^{\mu}{}_{\nu\rho}$ which we describe in the Supplement. Under a local $SO(2)$ rotation $\theta$ we have $\omega_{\mu} \to \omega_{\mu} + \partial_{\mu}\theta$, and in general there is nonzero torsion as determined by the Cartan structural equations.

The spatial curvature is related to $\omega$ as follows. The curvature constructed from $\omega$ is $d\omega$. On a constant-time, or spatial, slice $\Sigma$ with scalar curvature $R$ we have
\be
\int_{\Sigma} d\omega = \frac{1}{2}\int_{\Sigma} d^2x \sqrt{g} \, R\,.
\ee

The microscopic theory (and so also $S_{eff}$) is invariant under (i.) $U(1)$ gauge transformations, (ii.) coordinate reparameterizations, and (iii.) local $SO(2)$ rotations. The CS terms~\cite{Deser:1981wh} that can appear in $S_{eff}$ are then~\cite{bradlyn2014low}, in terms of differential forms,
\begin{align}
\begin{split}
	S_{CS} =&  \frac{\nu}{4\pi}\int_{\mathcal{M}} A \wedge dA 
	+ 2 \bar{s} A \wedge d\omega + \overline{s^2} \omega \wedge d\omega
  \label{E:CS} \\
	& + \frac{c}{96\pi}\int_{\mathcal{M}}I_{CS}[\Gamma]\,,
\end{split}
\end{align}
where with $\Gamma^{\mu}{}_{\nu} \equiv \Gamma^{\mu}{}_{\nu\rho}dx^{\rho}$ we have
\be
	I_{CS}[\Gamma] = \Gamma^{\mu}{}_{\nu} \wedge d\Gamma^{\nu}{}_{\mu} 
	+ \frac{2}{3}\Gamma^{\mu}{}_{\nu}\wedge \Gamma^{\nu}{}_{\rho}\wedge 
	\Gamma^{\rho}{}_{\mu}\,.
\ee
The second term in~\eqref{E:CS} is the WZ term, the third is sometimes called the second WZ term, and the last as the gravitational Chern-Simons (gCS) term.

 The dimensionless coefficients $(\nu,\bar{s},\overline{s^2},c)$ are known as the ``filling factor'', mean orbital spin per particle, mean orbital spin squared per particle, and chiral central charge. The flat-space Hall conductance is $\sigma_H = \frac{\nu}{2\pi}$, and when the space has curvature $R$, the Hall viscosity is $\eta_H = \frac{\bar{s}}{2}\rho+(12 \nu\, \text{var}(s) - c)\frac{R}{96\pi}$, with $\rho$ the charge density and $\text{var}(s)\equiv \overline{s^2}-\bar s^2$ the orbital spin variance~\cite{GCYFA}~\footnote{The combination $\text{var}(s)$ has been found to vanish for ``conformal block states''~\cite{bradlyn2015topological}.}.

 The third and fourth terms in~\eqref{E:CS} are related as
\be\la{wind}
	2\omega \wedge d\omega+I_{CS}[\Gamma] = \frac{1}{3}(\beta d\beta^{-1})^3\,,
\ee
where $\beta^{\mu}_a$ is the frame. The integral of the RHS of Eq.~\eqref{wind} over a closed space-time is proportional to an integer, a ``winding number'' of the frame over $\mathcal{M}$, so $\overline{s^2}$ and $c$ contribute to the bulk response only through the combination $12\nu \overline{s^2}-c$, or equivalently through $12\nu\, \text{var}(s)-c$. This combination and $(\nu,\bar{s})$ have been computed for integer quantum Hall states in~\cite{Abanov-2014, douglas2010bergman} and for various model fractional quantum Hall states in~\cite{can2014field, can2014fractional, can2014geometry, 2014-ChoYouFradkin, GCYFA, bradlyn2015topological, Klevtsov-fields, klevtsov2015precise}.

When the space has a boundary, $\text{var}(s)$ and $c$ can be disentangled. For example, it has been conjectured that the thermal Hall conductance of a quantum Hall state with an edge is given by $\kappa_H = c\frac{\pi}{3} k_B T$~\cite{2000-ReadGreen}. A similar relation has been shown to hold in any two-dimensional relativistic theory~\cite{Jensen:2012kj}. If this conjecture is correct, then measuring $\kappa_H$ would determine $c$, and $\text{var}(s)$ could be deduced from the Hall viscosity. 
 
\paragraph{Boundary terms and anomalies.} 

The CS terms in~\eqref{E:CS} are no longer invariant when $\mathcal{M}$ has boundary, leaving two possibilities for each CS term: (i) it cannot be made invariant by adding local boundary terms built from the external fields, or (ii.) it can. In the first case, we say that the CS term corresponds to an anomaly of a gapless edge theory, whose anomaly cancels the non-invariance of the bulk CS term via anomaly inflow. In the second case, the CS term does not correspond to an anomaly, and so does not protect the existence of gapless edge modes.

As we reviewed, the electromagnetic CS term (the first term in~\eqref{E:CS}) belongs to type (i.). Similarly, in relativistic field theories the gCS term is known to correspond to a boundary diffeomorphism anomaly~\cite{alvarez1985anomalies}. We have shown that in the non-relativistic setup relevant for this work, it is also impossible to construct local boundary terms canceling the diffeomorphism non-invariance of the gCS term and, therefore it corresponds to a diffeomorphism anomaly on the edge. This leaves the WZ terms.

To proceed, we describe the spacetime boundary $\partial\mathcal{M}$ via embedding functions $X^{\mu} = X^{\mu}(\sigma^\alpha)$ where $\mu=0,1,2$ and $(\sigma^{0},\sigma^{1})$ are boundary coordinates. The partial derivatives $\partial_\alpha X^{\mu}$ are tensors under both reparameterizations of the $x^{\mu}$ and the $\sigma^\alpha$. Using the $\partial_{\alpha}X^{\mu}$ and the bulk data $(\beta^{\mu}_a,\omega_{\mu})$, we can define a covariant derivative and the extrinsic curvature of the boundary. See the Supplement for the details. 

To illustrate the basic idea, consider the more familiar case with a time-dependent spatial metric $g_{ij}$. We consider  spatial boundaries whose shape does not change in time. Such a boundary can be parameterized as  $X^0=\sigma^0, X^i=X^i(\sigma^1)$. Given the $X^i$ one can construct tangent and normal vectors $t^i$ and $n^i$ that satisfy
\be
	n^i n_i = t^i t_i = 1\, , \qquad n_i t^i =0\,.
\ee

From this data we can construct an extrinsic curvature one-form $K_\alpha$ as
\be
\label{E:defK}
K_{\alpha} = n_i D_{\alpha} t^i\,.
\ee
The one-form $K_\alpha$ can be shown to be related to the spin connection projected to the boundary as
\be
\label{E:fromSpinToK}
	\omega_\alpha + K_\alpha = \partial_\alpha \varphi\,,
\ee
for a locally defined function $\varphi$. That is, the extrinsic curvature one-form differs from the spin connection (projected to the boundary) by an $SO(2)$ gauge transformation with boundary value $\varphi$.

Integrating over a spatial slice $\Sigma$ and using Stokes' theorem we obtain the Gauss-Bonnet theorem 
\be
	\frac{1}{2\pi}\left(\int_{\Sigma} d\omega +\int_{\partial\Sigma}K\right) = \chi\,,
\ee
where $\chi$ is the Euler characteristic of $\Sigma$, which is also the integer-valued winding number of $\varphi$ around $\partial\Sigma$.

The crucial point now is that we can use the extrinsic curvature $K_\alpha$ to render the WZ terms invariant by adding 
\be
 \label{E:WZbdy}
	S_{WZ,bdy} = \frac{\nu}{4\pi}\int_{\partial\mathcal{M}}
	\left( 2\bar{s} A \wedge K + \overline{s^2}\omega \wedge K\right)\,,
\ee
to the effective action. Equivalently, the contributions to effective action
\begin{align}
\label{E:WZ1}
	S_{WZ,1}& \equiv \frac{\nu\bar{s}}{2\pi} 
	\left(
	\int_{\mathcal{M}} A \wedge d\omega 
	+\int_{\mathcal{\p M}} A \wedge K
	\right)\,,
	\\
\label{E:WZ2}
	S_{WZ,2}& \equiv \frac{\nu\overline{s^2} }{4\pi}
	\left(
	\int_{\mathcal{M}} \omega \wedge d\omega
	+\int_{\mathcal{\p M}} \omega \wedge K
	\right)\,,
\end{align}
are invariant with respect to all symmetries of the problem, do not correspond to edge anomalies, and do not  necessitate gapless edge modes~\footnote{This statement was anticipated in~\cite{bradlyn2015topological}.}. This is the main result of this Letter. 

Putting the pieces together, we can write the total effective action as a sum
\be
\label{E:SeffDecompose}
S_{eff} = S_{CS}' + S_{WZ,1}+S_{WZ,2} + S_{edge} + \hdots\,,
\ee
where we have redefined the CS part of the action to only contain the terms that correspond to edge anomalies,
\be
\label{E:CS2}
S_{CS}' = \frac{\nu}{4\pi}\int_{\mathcal{M}}A \wedge dA + \frac{c}{96\pi}\int_{\mathcal{M}} I_{CS}[\Gamma] \,,
\ee
and the dots refer to additional, invariant bulk terms built from the external fields. The CS and gCS terms in~\eqref{E:CS2} protect the existence of a gapless edge theory $S_{edge}$, which varies under gauge transformations and infinitesimal reparameterizations $\xi^{\mu}$ as
\be
\label{E:totalInflow}
\delta S_{edge} = - \frac{\nu}{4\pi}\int_{\partial\mathcal{M}} \Lambda \, F - \frac{c}{96\pi}\int_{\partial\mathcal{M}} \partial_{\mu} \xi^{\nu} d\Gamma^{\mu}{}_{\nu}\,.
\ee

\paragraph{Lorentz and Galilean invariance.} 

Here we comment on the relation of this work to the literature. We regard the boundary term~\eqref{E:WZbdy} in a way which mirrors the situation in relativistic Hall states as discussed in~\cite{golkar2014euler}. The Riemann curvature can be dualized to the topologically conserved current $\mathcal{R}^{\mu} = \varepsilon^{\mu\nu\rho}\partial_{\nu}\omega_{\rho}$. $\mathcal{R}^{\mu}$ is the ``Euler current,'' in that its density is proportional to the Euler density $R$ on a spatial slice. The WZ term is just a coupling of $A_{\mu}$ to this conserved current. On a closed space, the ``charge'' associated with $\mathcal{R}^{\mu}$ is just the Euler characteristic of the spatial slice, and the conservation of $\mathcal{R}^{\mu}$ corresponds to the fact that this characteristic is a topological invariant which does not vary in time. On a space with boundary, the Euler characteristic includes an extrinsic boundary term, and so charge conservation mandates that the $A_{\mu}\mathcal{R}^{\mu}$ coupling must be supplemented with the extrinsic coupling in~\eqref{E:WZ1}.

The relativistic version of the WZ term was found in~\cite{golkar2014euler}. One can often obtain a Galilean-invariant theory from a relativistic one by taking a large speed of light limit as in~\cite{2006-SonWingate}. Taking this limit covariantly~\cite{Jensen:2014wha}, one gets a Galilean theory coupled to Newton-Cartan (NC) geometry (see e.g.~\cite{son-torsion,Jensen:2014aia}). Presumably the limit of the relativistic WZ term leads to the full WZ term~\eqref{E:WZ1} (modified to reflect Galilean invariance) \cite{moroz2015effective}. The relationship between edge physics and Hall viscosity in Galilean-invariant Hall states has also been discussed in~\cite{moroz2015galilean}.

\paragraph{Response.}

The CS~\eqref{E:CS2} and WZ terms~\eqref{E:WZ1},~\eqref{E:WZ2} lead to certain response functions which are protected by the symmetries as we now discuss.

Because $S_{edge}$ is an a priori unknown, gapless theory, we cannot completely fix the boundary response by the symmetries alone. We proceed by defining correlators of the $U(1)$ current $j^{\mu}$, spin current $s^{\mu}$, ``stress tensor'' $T^A_{\mu}$, and what we call the displacement operator $\mathcal{D}_{\mu}$. These are given by functional variations of $S_{eff}$ with respect to $(A_{\mu},\omega_{\mu},\beta^{\mu}_A,X^{\mu})$ respectively~\footnote{Note that a variation of $\omega_{\mu}$ at fixed $\beta^{\mu}_A$ is a variation of spatial torsion at fixed spatial metric.}. The symmetries imply that the displacement operator is along the normal vector $n^i$, and from it we find the external force density $\mathscr{F} = n^i \mathcal{D}_i$ which is required to fix the boundary.

The $U(1)$ current, spin current, and ``stress tensor'' have bulk and boundary components. For example, keeping $(\omega_{\mu},\beta^{\mu}_A)$ fixed, $j^{\mu}$ and $\mathcal{D}_{\mu}$ are defined via
\begin{align}
	\delta S_{eff} =& \int_{\mathcal{M}} [d^3x] \,\delta A_{\mu} j_{bulk}^{\mu} 
 \la{variation} \\
 \nonumber
	& \qquad+ \int_{\partial\mathcal{M}}[d^2\sigma]
	\left( \delta A_{\mu}j_{bdy}^{\mu} - \delta X^{\mu} \mathcal{D}_{\mu}\right)\,,
\end{align}
with $[d^3x]=d^3x \sqrt{g}$ and $[d^2\sigma]$ respectively an invariant bulk volume and boundary area. In other words, the current density is given by
\be
	j^{\mu} = j^{\mu}_{bulk} + j^{\mu}_{bdy}\delta(x^{\perp})\,,
\ee
with $\delta(x^{\perp})$ a delta function with support on $\partial\mathcal{M}$.  In principle, the boundary term in $\delta S_{eff}$ contains additional terms involving normal derivatives of $\delta A_{\mu}$. Those terms are not relevant for the rest of this Section.

All low-energy response functions of these operators are contained in $S_{eff}$. For illustrative purposes, we focus on the total charge $\mathscr{Q}$, and the contribution of the WZ terms~\eqref{E:WZ1},~\eqref{E:WZ2} to the total spin $\mathscr{S}$ and force density $\mathscr{F}$ exerted on the boundary. We consider a time-independent state in which the space is curved and threaded with magnetic flux.

The total charge is $\mathscr{Q} = \int_{\Sigma} d^2x \sqrt{g}\, j^0$, with $\Sigma$ a spatial slice. From $S_{eff}$ we find from~\eqref{E:SeffDecompose}
\begin{align}
	\begin{split}
	\label{E:Q}
		\mathscr{Q} = &\frac{\nu}{2\pi}\int_{\Sigma} F 
		+\frac{\nu \bar{s}}{2\pi}\left(\int_{\Sigma}d\omega
		+\int_{\partial\Sigma} K\right) +\mathscr{Q}_{edge}
	\\
		=&  \nu N_{\Phi} + \nu \bar{s}\chi + \mathscr{Q}_{edge}\,,
	\end{split}
\end{align}
where $N_{\Phi}$ and $\chi$ are the magnetic flux through and Euler characteristic of $\Sigma$, and $\mathscr{Q}_{edge}$ is the total charge coming from the edge theory~\footnote{More precisely, $\mathscr{Q}_{edge}$ is the gauge-invariant edge charge, which receives contributions both from $S_{edge}$ and~\eqref{E:AdA}.}. Here we have used that the local, gauge-invariant terms in the ellipsis of~\eqref{E:SeffDecompose} do not contribute to the total charge.

On a closed space,~\eqref{E:Q} becomes $\mathscr{Q} = \nu N_{\Phi} +\frac{ \nu \bar{s}}{2\pi} \int_{\Sigma}d\omega=\nu N_{\Phi} + \nu \bar{s} \chi$. This expression was already known in the FQH literature \cite{haldane1983fractional,WenZeeShiftPaper}. Eq.  \eqref{E:Q} generalizes it to systems with an edge. The effect of the boundary term~\eqref{E:WZbdy} is to ensure that there is an extrinsic contribution to $\mathscr{Q}$ in such a way that the total charge depends on $\bar{s}$ only through the Euler characteristic $\chi$ of the spatial slice. 

The total spin $\mathscr{S} = \int_{\Sigma} d^2x\sqrt{g} \,s^0$ is
\be
\label{E:S}
	\mathscr{S} = \nu \bar{s} N_{\Phi}
	+ \nu \overline{s^2}\chi+\hdots\,.
\ee
The dots indicate contributions from the rest of $S_{eff}$, including the gCS term. A similar relation has appeared in \cite{hughes2012torsion} when space-time is compact. The boundary term~\eqref{E:WZbdy} gives an extrinsic contribution to $\mathscr{S}$, ensuring that it depends on $\overline{s^2}$ only through $\chi$.

Finally, the external force density $\mathscr{F}=n^i\mathcal{D}_i$ as
\be
\mathscr{F} =- \frac{\nu\bar{s}}{2\pi}\left( t^i\partial_i E_{||} + K E_{\perp}\right) - \frac{\nu\overline{s^2}}{4\pi}\left( t^i \partial_i \mathcal{E}_{||} + K \mathcal{E}_{\perp}\right)+\hdots\,,
\ee
where again the dots indicate contributions from the rest of $S_{eff}$. Here $E_{||} $ and $E_{\perp}$ the electric fields parallel and normal to the boundary (and similarly for the components of ``gravi-electric'' field $\mathcal{E}_{i}=\p_{0} \omega_{i}-\p_{i}\omega_{0}$),
and $K=t^iK_i$ the geodesic curvature of the boundary.

\paragraph{Relation to index theorem.}

There is an intimate connection between quantum anomalies in relativistic field theory and index theorems \cite{alvarez1985structure}. It is natural to ask if there is any connection between Hall states and index theorems for manifolds with boundary. Here we illustrate such a connection in the simplest case of non-interacting electrons. Namely, we assume that we have $\mathscr{Q}$ non-interacting electrons and  (i) only the lowest Landau level (LLL) is filled and (ii) we apply particular boundary conditions for the bulk electrons. In this system, $\nu=1$ and $\bar{s}=\frac{1}{2}$, and the LLL states are zero modes of the anti-holomorphic differential operator of momentum $\bar{ D}$ on the spatial slice. The number of such zero modes is counted by the Atiyah-Patodi-Singer (APS) index theorem~\cite{atiyah1975spectral} provided that the electrons obey so-called APS boundary conditions. The index of $\bar{D}$ is
\be
 \label{E:index}
	\text{ind}(\bar{D}) = N_{\Phi} + \frac{1}{2}\chi + \frac{1}{2}\eta\,,
\ee
where $N_{\Phi}$ and $\chi$ are as above, the ``$\eta$-invariant'' is
\be
	\eta\equiv \text{sign}\,\bar{D}|_{\partial\Sigma} = \sum \text{sign} \, \lambda\,,
\ee
where $\bar{D}|_{\partial\Sigma}$ is $\bar{D}$ restricted to the boundary, and the sum runs over eigenmodes of this operator with eigenvalue $\lambda$~\cite{eguchi1980gravitation}. Note that the index~\eqref{E:index} indeed matches our general expression~\eqref{E:Q} for $\nu = 1, \bar{s}=\frac{1}{2}$, and $\mathscr{Q}_{edge} = \frac{\eta}{2}$. 

The total number of electrons $\mathscr{Q}$ is integer, which is guaranteed in~\eqref{E:index} by the $\eta$-invariant. For example, if the spatial slice is a disk $\chi=1$, then $\eta=1 - 2\{ N_\Phi\}$, where $\{N_{\Phi}\}$ is the non-integer part of $N_{\Phi}$. Then $\text{ind}(\bar{D})=\mathscr{Q}=\lc N_\Phi\rc+1$, indeed giving integer $\mathscr{Q}$. 

\paragraph{Singular expansion of charge density.}

So far our results have been obtained only from the symmetries of the problem. As an application, we derive the singular expansion of the charge density of a flat-space Hall state. From $S_{eff}$ we obtain the charge density $\rho=j^0$
\be
	\rho= \frac{\nu B}{2\pi}\theta(\Sigma) +\left(\frac{\nu \bar{s}}{2\pi}K+j^0_{bdy}\right)\delta(\p\Sigma) 
	+\frac{\zeta}{2\pi}\p_n\delta(\p\Sigma) +\ldots\,.
 \la{singexp}
\ee
Here $\p_n\delta(\p\Sigma)$ denotes the normal derivative of the delta function on the boundary of the system. The first term of~\eqref{singexp} comes from~\eqref{E:AdA}, the second from the boundary part of the first WZ term~\eqref{E:WZ1} and $j^0_{bdy}$ (defined in~\eqref{variation}) depends on the non-universal details of $S_{edge}$. The third comes from two invariant, higher order terms in $S_{eff}$,
\be
 	\frac{\sigma_H^{(2)}}{2\pi}\int_{\mathcal{M}}[d^3x]\,B D^iE_i \, ,
 	\quad \frac{\xi}{2\pi} \int_{\p \mathcal{M}}[d^2\sigma] \,n^iE_i\, ,
\ee
with $\zeta = \sigma_H^{(2)}+\xi$. Here $\sigma_H^{(2)}$ is the $O(k^2)$ correction to the Hall conductivity, and $\xi$ is a dimensionless parameter related to the total dipole moment at the edge. The coefficient $\zeta$ is relevant for the so-called ``overshoot'' phenomenon \cite{can2014singular} and for the Laughlin function is related to the Hall viscosity. When the underlying system is Galilean-invariant, $\sigma_H^{(2)}$ gets a contribution from the Hall viscosity~\cite{2012-HoyosSon}, thus relating the ``overshoot'' with $\eta_H$.

For simplicity we take $\Sigma$ to be a flat disk of radius $R$. Then~\eqref{singexp} becomes
\begin{align}
\begin{split}
\label{lrho}
	\rho  =&\frac{\nu B}{2\pi} \Theta(R-r) +\left(\frac{\nu\bar{s}}{\pi} +2Rj^0_{bdy}\right)\delta(r^2-R^2) 
 \\
 & \qquad  + \frac{\zeta}{2\pi}R^2\delta'(r^2-R^2)+\hdots\,.
\end{split}
\end{align}
Specifying for Laughlin's state with $\nu= \frac{1}{2n+1}$ and $\nu \bar{s} = \frac{1}{2}$, this matches the singular expansion obtained by Wiegmann and Zabrodin~\cite{2006-ZabrodinWiegmann} directly from the Laughlin's wave function for $\zeta=1-2\nu$ and $j^0_{bdy}=-\frac{\nu \bar{s}}{2\pi R}$~\footnote{The value of $\zeta$ depends on the boundary conditions of the problem. For Laughlin's droplet made out of $N$ particles on an infinite plane one can obtain $\zeta$ by matching~\eqref{lrho} to the exact sum rule \protect{$\langle \sum_{i=1}^N |z_i|^2\rangle = l^2N(\nu^{-1}N+2-\nu^{-1})$}.}. One can also match for an infinitesimally different definition of the radius $R$, in which case $\zeta$ is unchanged but $j^0_{bdy}=0$.

\paragraph{Conclusions.}

Using effective field theory and symmetries on a space with boundary, we have made a systematic study of the Chern-Simons terms~\eqref{E:CS} that appear in the low-energy effective action of quantum Hall states.

The main result is that the WZ terms are not Chern-Simons terms per se, but rather the couplings of $A_{\mu}$ and the spin connection $\omega_\mu$ to a topologically conserved but non-trivial ``Euler current.'' On a space with boundary, these bulk couplings must be supplemented with boundary couplings between $A_{\mu}$ and the spin connection $\omega_\mu$ to the extrinsic curvature of the edge.

An immediate corollary to our result is that the coefficients of the WZ terms, $\bar{s}$ and $\overline{s^2}$, can jump across an interface without closing a gap or breaking the symmetries of the problem, namely $U(1)$ gauge invariance, coordinate reparameterizations, or local $SO(2)$ invariance.

Our work suggests several open questions. One regards the status of the CS terms and boundary physics in an approximately Galilean-invariant Hall state, where the electromagnetic CS term~\eqref{E:AdA} is $U(1)$ but not boost invariant (see e.g.~\cite{2012-HoyosSon}). In such a state, do the WZ terms correspond to boundary terms as here? More generally, what are the symmetry protected topological phases with Galilean symmetry?

We are pleased to thank T.~Can, A.~Cappelli, S.~ Golkar, G.~Monteiro, A.~Kapustin and P. Wiegmann for useful discussions and comments. This work was supported in part by the NSF under grants PHY-0969739 and DMR-1206790. 

Note added: After this work was completed, the authors of~\cite{moroz2015galilean} have privately informed us that the results of this Letter are consistent with theirs.

\bibliography{Bibliography}





\end{document}


\title{Supplementary material for: Boundary effective action for quantum Hall states}

\author{Andrey~Gromov}
\affiliation{Department of Physics and Astronomy, Stony Brook University,  Stony Brook, NY 11794, USA}

\author{Kristan~Jensen}
\affiliation{C.N.~Yang Institute for Theoretical Physics, SUNY Stony Brook,  Stony Brook, NY 11794, USA}

\author{Alexander G.~Abanov}
\affiliation{Department of Physics and Astronomy, Stony Brook University,  Stony Brook, NY 11794, USA}
\affiliation{Simons Center for Geometry and Physics, Stony Brook University,  Stony Brook, NY 11794, USA}

\date{\today}


\maketitle

\section{ Preliminary comments}

In the main body, we began our primary analysis by coupling field theories with a spatial stress tensor $T^{ij}$ to an external spatial metric $g_{ij}$. To linear order in fluctuations $h_{ij}$ of $g_{ij}$ around flat space, $g_{ij} = \delta_{ij} + h_{ij}$, the appearance of $h_{ij}$ in $S_{eff}$ is fixed to be
\be
\label{E:linear}
S_{eff} [g]= S_{eff}[\delta] +\frac{1}{2} \int_{\mathcal{M}} dtd^dx \, h_{ij} T^{ij} + \mathcal{O}(h^2)\,.
\ee
In principle, this is enough information to compute correlation functions of $T^{ij}$ and other operators at nonzero separation in flat space. However, in many applications it is useful to understand the coincident limit. For example in a gapped phase all correlation functions are approximately local on length scales longer than the correlation length. To discuss the coincident limit of correlation functions of $T^{ij}$, we need to specify a prescription for the $\mathcal{O}(h^2)$ and higher terms in $S_{eff}$. Different prescriptions, much like different regulatory schemes in field theory, can be chosen to preserve different symmetries. In this work we implicitly choose for the nonlinear couplings of $h$ to respect coordinate reparameterizations. For example, consider the theory of a non-relativistic complex field $\Psi$ whose flat-space action is
\be
\label{E:Sfree}
S_{free} = \int dt d^dx \left\{ \frac{i}{2}\Psi^{\dagger} \overleftrightarrow{\partial}_0 \Psi - \frac{\delta^{ij}}{2m}\partial_i \Psi^{\dagger}\partial_j \Psi\right\}\,.
\ee
This theory can be coupled to $g_{ij}$ in a way that respects spatial reparameterizations by changing it to
\be
S_{free} \to \int dt d^dx \sqrt{g} \left\{ \frac{i}{2}\Psi^{\dagger}\overleftrightarrow{\partial}_0 \Psi - \frac{g^{ij}}{2m}\partial_i \Psi^{\dagger}\partial_j \Psi\right\}\,.
\ee

This prescription is not enough to fully specify the curved-space theory. For example, 
\begin{align}
\begin{split}
\int dt d^dx \sqrt{g}& \left\{ \frac{i}{2}\Psi^{\dagger}\overleftrightarrow{\partial}_0 \Psi - \frac{g^{ij}}{2m}\partial_i \Psi^{\dagger}\partial_j \Psi \right.
\\
&\left. \qquad - \frac{\alpha}{2m} R |\Psi|^2\right\}\,,
\end{split}
\end{align}
with $R$ the scalar curvature of $g_{ij}$, is invariant under spatial reparameterizations for any value of $\alpha$. $\alpha$ is a coupling of the curved-space theory. In defining the curved-space theory, we not only demand that the action is invariant under spatial reparameterizations, but we must also specify all of the curved-space couplings.

Observe that, by construction, the curved space action is now invariant under an infinite-dimensional family of coordinate transformations. So far this is a statement about classical field theory, but it often survives quantum corrections. The full partition function will also be invariant under coordinate transformations, up to a possible quantum anomaly.

This symmetry -- the invariance under the theory under spatial reparameterizations -- is a ``spurionic symmetry'' in the language of high energy physics. To explain this term, we regard $g_{ij}$ as a coupling of the quantum theory. Under infinitesimal coordinate transformations $x^i \to x^i + \xi^i$, $g_{ij}$ is not invariant but instead transforms as 
\be
\delta_{\xi} g_{ij} =\xi^k \partial_k g_{ij} + g_{ik}\partial_j \xi^k + g_{jk} \partial_i \xi^k\,.
\ee
So a coordinate transformation leaves the action invariant, but the couplings of the theory transform. This is the meaning of a spurionic symmetry.

A theory with a global $U(1)$ symmetry, coupled to a background electromagnetic field $A_{\mu}$, also possesses a spurionic symmetry under which $A_{\mu}$ (which we regard as a coupling of the theory) transforms as $A_{\mu} \to A_{\mu} + \partial_{\mu} \Lambda$.

In some sense, spurionic symmetries are trivial. In the case of spatial reparameterizations, one can always begin with an ordinary flat space theory and tune its nonlinear couplings to $g_{ij}$ to make it invariant. Yet spurionic symmetries are rather useful, as they constrain the full partition function of the theory.

Global symmetries are a subset of spurionic ones. A global symmetry is a particular spurionic symmetry under which all of the couplings are invariant. For example, if our theory is in flat space $g_{ij} = \delta_{ij}$, and all other nonzero couplings are constant scalars, then the global symmetries include translations and rotations, under which $g_{ij}$ and the other couplings are invariant. Noether's theorem applies to continuous global symmetries, not spurionic ones: using the transformation that generates the global symmetry, one can construct a conserved Noether current operator.

With all of this in mind, it should not be a surprise that we can do better. We can start with a flat-space theory and tune its couplings to external fields so as to make it invariant under an arbitrary change of coordinates, which depend on both space and time.

Let us see how this works for the free field theory~\eqref{E:Sfree}. It is clear what we need to do: we replace $\partial_0$ with $v^{\mu}\partial_{\mu}$, where $v^{\mu}$ is a nowhere-vanishing vector field, and replace $\delta^{ij}$ with a rank-$d$ semi-positive, symmetric tensor $g^{\mu\nu}$. We also demand that $v^{\mu}v^{\nu}+g^{\mu\nu}$ is non-degenerate. The fields $(v^{\mu},g^{\mu\nu})$ are the external fields, which transform as tensors under an arbitrary coordinate transformation. Letting $\Psi$ transform as a scalar, the functional
\be
\int d^{d+1}x \sqrt{\gamma} \left\{ \frac{iv^{\mu}}{2}\Psi^{\dagger}\overleftrightarrow{\partial}_{\mu}\Psi - \frac{g^{\mu\nu}}{2m}\partial_{\mu}\Psi^{\dagger}\partial_{\nu}\Psi\right\}\,,
\ee
with $\sqrt{\gamma}$ a good measure defined below, is a curved version of~\eqref{E:Sfree} invariant under any coordinate transformation. As above, this statement often survives quantum corrections. 

The external fields $(v^{\mu},g^{\mu\nu})$ can be understood as describing some ``geometry.'' To get a sense for it, we can locally choose coordinates where $v^{\mu}=\delta^{\mu}_t$. If we pick $g^{0\mu}=0$, then the nonzero components of $g$ are $g^{ij}$ which gives an inverse spatial metric on slices of constant time. This ``geometry'' is a version of what is known as Newton-Cartan (NC) geometry. Note that it automatically appears if we write the theory of a non-relativistic free field~\eqref{E:Sfree} in a coordinate-free way.

We require some details of this geometry, including definitions for a covariant derivative and the extrinsic curvature of a boundary.

\section{Newton-Cartan geometry in the bulk}

We continue with NC geometry on a $(d+1)$-dimensional, orientable spacetime $\mathcal{M}$ without boundary. There are different versions of NC geometry. Much ink~\cite{Christensen-2014,son2013newton,son-torsion,Jensen:2014aia,Geracie:2015dea}  has been spilled lately on a version which naturally arises in the context of Galilean field theories. We will not use this version, but instead stick with one which gives a set of sources which naturally couple to a non-relativistic, non-Galilean field theory.

The version we require is formulated nicely in~\cite{bradlyn2014low}. Here we summarize the basic data which we need to define extrinsic geometry in the next Appendix, as well as some differential geometry which is useful to keep in one's back pocket.

One parameterization is in terms of a basis of tangent vectors $\beta_a^{\mu}$, with $a=0,1,..,d$, their inverse $(\beta^{-1})^a_{\mu}$, and a spin connection $\omega^a{}_{b\mu}$. The $\beta_a^{\mu}$ give a local choice of frame, and $\beta^{-1}$ a ``coframe.'' All of these objects are genuine tensors under coordinate reparameterizations. We continue by separating the frame and coframe into a time (co)vector and a basis of spatial (co)vectors, denoting
\begin{align}
\begin{split}
v^{\mu} &\equiv \beta_0^{\mu} \,,\qquad n_{\mu} \equiv (\beta^{-1})^0_{\mu}\,,
\\
 E_A^{\mu} &\equiv \beta_A^{\mu} \,, \qquad e^A_{\mu} \equiv (\beta^{-1})^A_{\mu}\,,
\end{split}
\end{align}
where $A,B=1,..,d$ index the basis of spatial (co)vectors. We restrict the spin connection to only have antisymmetric spatial components,
\be
\omega^0{}_{a\mu} = 0\,, \qquad \omega^a{}_{0\mu} = 0\,, \qquad \omega^{(AB)}_{\mu} = 0\,,
\ee
where in the last expression we have raised the second index with $\delta^{AB}$, and round brackets denote symmetrization. From the spatial frame and coframe we obtain
\be
g^{\mu\nu} \equiv E_A^{\mu}E^{\nu}_B \delta^{AB}\,, \qquad g_{\mu\nu} \equiv e^A_{\mu}e^B_{\nu} \delta_{AB}\,.
\ee
$g_{\mu\nu}$ is the covariant version of a spatial metric $g_{ij}$, and $g^{\mu\nu}$ the covariant version of its inverse $g^{ij}$. Note that
\begin{align}
\begin{split}
v^{\mu}n_{\mu}&=1\,, \quad g_{\mu\nu}v^{\nu}=0\,, \\
 g^{\mu\nu}n_{\nu}&=0\,, \quad g^{\mu\rho}g_{\nu\rho} = \delta^{\mu}_{\nu} - v^{\mu}n_{\nu}\,.
\end{split}
\end{align}
Further, $(v^{\mu},g^{\mu\nu})$ are determined algebraically from $(n_{\mu},g_{\mu\nu})$ and vice versa. By construction
\be
\label{E:gamma}
\gamma_{\mu\nu} \equiv n_{\mu}n_{\nu} + g_{\mu\nu}\,,
\ee
is a positive tensor from which we can define a covariant integration measure, $d^{d+1}x \sqrt{\gamma}$. We can also define a epsilon tensor via
\be
\varepsilon^{\mu_1\hdots \mu_{d+1}} \equiv \frac{\epsilon^{\mu_1\hdots \mu_{d+1}}}{\sqrt{\gamma}}\,,
\ee
where $\epsilon^{\mu_1\hdots \mu_{d+1}}$ is an epsilon symbol with $\epsilon^{t1\hdots d} = +1$.

From the frame and spin connection we can define an ordinary connection $\Gamma^{\mu}{}_{\nu\rho}$, which is an NC analogue of the Levi-Civita connection of Riemannian geometry. There are in fact many different connections $\Gamma$ that can be defined from the tensor data at hand. The one we use is
\be
\label{E:Gamma}
\Gamma^{\mu}{}_{\nu\rho}  = \beta^{\mu}_a \partial_{\rho} (\beta^{-1})^a_{\nu} + \beta^{\mu}_a \omega^a{}_{b\rho} (\beta^{-1})^b_{\nu}\,,
\ee
so that
\be
\partial_{\mu}\beta^{\nu}_a + \Gamma^{\nu}{}_{\rho\mu}\beta^{\rho}_a - \beta^{\nu}_b \omega^b{}_{a\mu}=0\,.
\ee
The covariant derivative $D_{\mu}$ of a tensor, say a mixed tensor $\mathfrak{T}^{\nu}{}_{\rho}$, is given in terms of $\Gamma$ via
\be
D_{\mu} \mathfrak{T}^{\nu}{}_{\rho} = \partial_{\mu} \mathfrak{T}^{\nu}{}_{\rho} + \Gamma^{\nu}{}_{\sigma\mu} \mathfrak{T}^{\sigma}{}_{\rho} - \mathfrak{T}^{\nu}{}_{\sigma} \Gamma^{\sigma}{}_{\rho\mu}\,.
\ee

One can readily verify that $(n_{\mu},g_{\nu\rho})$ (and so also $(v^{\mu},g^{\nu\rho})$) are covariantly constant,
\be
D_{\mu}n_{\nu} = 0\,, \qquad D_{\mu}g_{\nu\rho} = 0\,.
\ee

We define the curvature $R^{\mu}{}_{\nu\rho\sigma}$ and torsion $T^{\mu}{}_{\nu\rho}$ from $\Gamma$ in the usual way. For $\mathfrak{T}^{\mu}{}_{\nu}$ a mixed tensor, the commutator of covariant derivatives is
\be
[D_{\rho},D_{\sigma}]\mathfrak{T}^{\mu}{}_{\nu} = R^{\mu}{}_{\alpha\rho\sigma}\mathfrak{T}^{\alpha}{}_{\nu} - \mathfrak{T}^{\mu}{}_{\alpha}R^{\alpha}{}_{\nu\rho\sigma} - T^{\alpha}_{\rho\sigma}D_{\alpha}\mathfrak{T}^{\mu}{}_{\nu}\,.
\ee
This definition is equivalent to the following. Let $\Gamma^{\mu}{}_{\nu} = \Gamma^{\mu}{}_{\nu\rho}dx^{\rho}$ be a one-form built from $\Gamma$. Then the curvature two-form $R^{\mu}{}_{\nu}$ is
\be
\label{E:Riemann}
R^{\mu}{}_{\nu} = d\Gamma^{\mu}{}_{\nu} + \Gamma^{\mu}{}_{\rho} \wedge \Gamma^{\rho}{}_{\nu} = \frac{1}{2}R^{\mu}{}_{\nu\rho\sigma}dx^{\rho}\wedge dx^{\sigma}\,,
\ee
and the torsion is
\be
T^{\mu}{}_{\nu\rho} = \Gamma^{\mu}{}_{\rho\nu} - \Gamma^{\mu}{}_{\nu\rho}\,. 
 \la{tordef}
\ee

Alternatively we could compute the curvature and torsion from the coframe and spin connection. Writing the coframe as a vector-valued one-form $(\beta^{-1})^a = (\beta^{-1})^a_{\mu}dx^{\mu}$ and the spin connection as a matrix-valued one-form, $\omega^a{}_b = \omega^a{}_{b\mu}dx^{\mu}$, the torsion is constructed from the coframe and spin connection to be
\be
T^a = d(\beta^{-1})^a + \omega^a{}_b \wedge (\beta^{-1})^b\,.
\ee
This is related to~\eqref{tordef} as
\be
T^{\mu}{}_{\nu\rho} = \beta^{\mu}_a T^a{}_{\nu\rho}\,.
\ee
Note that $T^{\mu}{}_{\nu\rho}$ is not arbitrary; from the definition above, one can show that it satisfies two constraints~\cite{bradlyn2014low}
\begin{align}
\begin{split}
\label{E:Torsion}
n_{\mu}T^{\mu}_{\nu\rho} &= \partial_{\nu}n_{\rho} - \partial_{\rho}n_{\nu}\,,
\\
(T_{\mu\nu\rho} + T_{\nu\mu\rho})v^{\rho}& = - \pounds_v g_{\mu\nu}\,,
\end{split}
\end{align}
where $\pounds_v$ indicates a Lie derivative along $v$ and we have lowered the first index of $T$ with $g_{\mu\nu}$.

The first condition in~\eqref{E:Torsion} implies that non-trivial $n$ mandates torsion. To understand the second, pick coordinates so that $v^{\mu} = \delta^{\mu}_t$, in which case $g_{\mu\nu}$ only has spatial components $g_{ij}$. The RHS of the second condition in~\eqref{E:Torsion} is
\be
T_{ijt} + T_{jit} = - \dot{g}_{ij}\,.
\ee
So a time-dependent spatial metric also mandates torsion.

The curvature of the spin connection is
\be
R^a{}_b = d\omega^a{}_b + \omega^a{}_c \wedge \omega^c{}_b = \frac{1}{2}R^a{}_{b\mu\nu}dx^{\mu} \wedge dx^{\nu}\,.
\ee
Since $\omega^a{}_b$ only has spatial components, so does $R^a{}_b$, i.e. its only nonzero components are $R^A{}_B$. Converting the $a,b$ indices of $R^a{}_b$ to spacetime indices through the frame, $R^a{}_b$ is equivalent to the Riemann curvature in~\eqref{E:Riemann}
 \be
R^{\mu}{}_{\nu\rho\sigma} = \beta_a^{\mu}(\beta^{-1})^b_{\nu} R^a{}_{b\rho\sigma}\,.
\ee

A straightforward computation shows that the $\Gamma$ in~\eqref{E:Gamma} is in fact determined by $(n_{\mu},h_{\mu\nu},T^{\mu}{}_{\nu\rho})$ (up to the constraints~\eqref{E:Torsion} on the torsion) as
\begin{align}
\begin{split}
\label{E:Gamma2}
\Gamma^{\mu}{}_{\nu\rho} = &v^{\mu} \partial_{(\nu}n_{\rho)} + \frac{1}{2}g^{\mu\sigma}\left( \partial_{\nu}g_{\rho\sigma} + \partial_{\rho}g_{\nu\sigma} - \partial_{\sigma} g_{\nu\rho}\right)
\\
& \qquad - \frac{1}{2}\left( T^{\mu}{}_{\nu\rho} - T_{\nu}{}^{\mu}{}_{\rho} + T_{\rho\nu}{}^{\mu}\right)\,,
\end{split}
\end{align}
where we have raised and lowered indices in the second line with $g_{\mu\nu}$ and $g^{\mu\nu}$.

The next, crucial step, is to introduce a transformation which amounts to invariance under local spatial rotations. We will then demand that field theories coupled to NC geometry are invariant under these local $SO(d)$ rotations, in the same way that we will demand invariance under coordinate reparameterizations. On the frame, these local rotations simply rotate the spatial vectors $E_A^{\mu}$ into each other. At the infintesimal level, we parameterize a local spatial rotation as $v^A{}_B$ with $v^{(AB)}=0$. The frame and coframe vary as
\begin{align}
\begin{split}
\delta_v v^{\mu}& = 0\,, \qquad \delta_v E_A^{\mu} =  E_B^{\mu}v^B{}_A\,, \\
\delta_v n_{\mu} & = 0\,, \qquad \delta_v e^A_{\mu} =- v^A{}_B e^B_{\mu}\,,
\end{split}
\end{align}
and the spin connection transforms as an $SO(d)$ connection,
\be
\delta_v \omega^A{}_{B\mu} = \partial_{\mu} v^A{}_B + \omega^A{}_{C\mu} v^C{}_B - v^A{}_C \omega^C{}_{B\mu}\,.
\ee
One can think of this local $SO(d)$ as a redundancy introduced when decomposing the spatial metric $g_{\mu\nu}$ into a basis of spatial covectors.

In mathematical parlance, we have used the data $(n_{\mu},g_{\nu\rho})$ to (locally) reduce the frame bundle $F\mathcal{M}$ from a $GL(d+1)$ bundle over $\mathcal{M}$ to an $SO(d)$ bundle. This procedure is globally defined only if $(n_{\mu},g_{\nu\rho})$ are globally defined and non-singular with $g$ everywhere of rank $d$.

The reader can readily verify that the simplest $SO(d)$-invariant objects are
\be
 n_{\mu}\,, \quad g_{\mu\nu}\,, \quad \Gamma^{\mu}{}_{\nu\rho}\,, 
\ee
and so also $(v^{\mu},g^{\nu\rho})$. Since the torsion and curvature are constructed from $\Gamma$,
\be
T^{\mu}{}_{\nu\rho}\,, \quad R^{\mu}{}_{\nu\rho\sigma}\,,
\ee
are $SO(d)$-invariant too. Indeed, using~\eqref{E:Gamma2}, we can specify all $SO(d)$-invariant data in terms of $(n_{\mu},g_{\nu\rho},T^{\mu}{}_{\nu\rho})$.

That is, we could also define this version of NC geometry from
\be
\label{E:AlternativeNC}
n_{\mu}\,, \qquad g_{\mu\nu}\,, \qquad T^{\mu}{}_{\nu\rho}\,,
\ee
from which one then reconstructs $(v^{\mu},g^{\mu\nu})$, provided that the torsion satisfies~\eqref{E:Torsion}. From $(n_{\mu},g_{\mu\nu})$ one can build a coframe $\beta^{-1}$ up to an $SO(d)$ redundancy. 

Both ways of thinking about this NC geometry -- in terms of a frame and $SO(d)$ spin connection, or in terms of the spacetime data in~\eqref{E:AlternativeNC} -- are complementary. It is helpful to switch from one presentation to the other depending on the problem at hand.

Now we specialize to $d=2$. Then the local $SO(d)$ redundancy is abelian, and the spin connection satisfies
\be
\omega^A{}_B = \varepsilon^A{}_B \,\omega\,,
\ee
where $\varepsilon^1{}_2 = +1$ is the covariantly constant epsilon tensor with spatial frame indices. Under a local $SO(2)$ rotation $v^A{}_B = \varepsilon^A{}_B \,v$, the abelianzed connection $\omega$ transforms as $\delta_v \omega = dv$. The Riemann curvature also simplifies as
\be
R^A{}_B = \varepsilon^A{}_B \, \mathcal{R}\,, \qquad \mathcal{R} = d\omega\,.
\ee
We also have
\be
\mathcal{R} = \frac{1}{2}\varepsilon^{\mu\nu\rho}n_{\mu}R_{\nu\rho}\,,
\ee
with $R_{\mu\nu}=g_{\mu\rho}R^{\rho}{}_{\nu}$ and $R^{\mu}{}_{\nu}$ the Riemann curvature form.

Finally, we introduce an exterior covariant derivative $D$ which will be useful in the next Appendix. $D$ is defined to act on forms which may also carry spacetime indices, and it takes a $p$-form with indices to a $p+1$-form of the same type. For example, on a matrix-valued $p$-form $U^{\mu}{}_{\nu}$, a vector-valued $m$-form $Y^{\mu}$, and a covector-valued $n$-form $Z_{\mu}$ it acts as
\begin{align}
\begin{split}
DU^{\mu}{}_{\nu} &= dU^{\mu}{}_{\nu} + \Gamma^{\mu}{}_{\rho} \wedge U^{\rho}{}_{\nu} - (-1)^{p}U^{\mu}{}_{\rho}\wedge \Gamma^{\rho}{}_{\nu}\,,
\\
DY^{\mu} & = dY^{\mu} + \Gamma^{\mu}{}_{\nu} \wedge Y^{\nu}\,, 
\\
 DZ_{\mu} &= dZ_{\mu} -(-1)^n Z_{\nu}\wedge \Gamma^{\nu}{}_{\mu}\,.
\end{split}
\end{align}
This operator is useful, satisfying
\begin{align}
\begin{split}
d(Y^{\mu}\wedge Z_{\mu}) &= DY^{\mu} \wedge Z_{\mu} + (-1)^m Y^{\mu}\wedge DZ_{\mu}\,,
\\
DR^{\mu}{}_{\nu} &= 0\,,
\end{split}
\end{align}
along with
\begin{align}
\begin{split}
D^2 U^{\mu}{}_{\nu} &= [R,U]^{\mu}{}_{\nu}\,, 
\\
D^2 Y^{\mu}& = R^{\mu}{}_{\nu}\wedge Y^{\nu}\,,
\\
D^2 Z_{\mu} & = - Z_{\nu}\wedge R^{\nu}{}_{\mu}\,.
\end{split}
\end{align}

\section{Newton-Cartan geometry on spaces with boundary}

Now we turn to study NC geometry on orientable spaces $\mathcal{M}$ with a boundary $\partial\mathcal{M}$. We describe the boundary covariantly via embedding functions $X^{\mu}(\sigma^{\alpha})$ where the $\sigma^{\alpha}$ are coordinates on $\partial\mathcal{M}$. The $X^{\mu}$ themselves are not tensors, but the $f_{\alpha}^{\mu} \equiv \partial_{\alpha}X^{\mu}$ are.

The $f_{\alpha}^{\mu}$ allow us to project any tensor on $\mathcal{M}$ with lower indices to a tensor on $\partial\mathcal{M}$. For example,
\be
n_{\alpha} = f_{\alpha}^{\mu}n_{\mu}\,.
\ee
That is, the $f^{\mu}_{\alpha}$ allow us to ``pullback'' covariant tensors on $\mathcal{M}$ to covariant tensors on $\partial\mathcal{M}$. We denote this operation as $\text{P}[h]$ for $h$ a covariant tensor, e.g.
\be
\text{P}[n] = n_{\alpha} d\sigma^{\alpha}\,.
\ee
Note that we can only pullback covariant tensors so far. We require a metric to ``pullback'' contravariant tensors.

In the previous Appendix we defined the positive tensor $\gamma_{\mu\nu}=n_{\mu}n_{\nu} + g_{\mu\nu}$, which can serve as a Riemannian metric on $\mathcal{M}$. We consider smooth boundaries so that $\text{P}[\gamma]$ is also a positive tensor $\gamma_{\alpha\beta}$, whose inverse we denote as $\gamma^{\alpha\beta}$.  Using $\gamma^{\alpha\beta}$ and $\gamma_{\mu\nu}$ we define
\be
f^{\alpha}_{\mu} \equiv \gamma^{\alpha\beta}\gamma_{\mu\nu} f_{\beta}^{\nu}\,.
\ee
The $f^{\alpha}_{\mu}$ allow us to project upper indices, inducing contravariant tensors on $\partial\mathcal{M}$ from contravariant tensors on $\mathcal{M}$, e.g.
\be
v^{\alpha} = f^{\alpha}_{\mu}v^{\mu}\,.
\ee

We have all the data required to build a  covector $N_{\mu}$ normal to $\partial\mathcal{M}$. From $\gamma_{\alpha\beta}$ we can also construct an epsilon tensor on $\partial\mathcal{M}$, $\varepsilon^{\alpha_1\hdots \alpha_d}$, from which we define
\be
N_{\mu} = \frac{1}{d!}\varepsilon_{\mu \nu_1 \hdots \nu_d}\varepsilon^{\alpha_1\hdots  \alpha_d} f_{\alpha_1}^{\nu_1}\hdots f_{\alpha_d}^{\nu_d}\,,
\ee
which is normal in the sense that
\be
N_{\alpha} = f_{\alpha}^{\mu}N_{\mu} = 0\,.
\ee
We also define $N^{\mu} = \gamma^{\mu\nu}N_{\nu}$, which conveniently satisfies
\be
N_{\mu}N^{\mu}=1\,.
\ee
Using $N_{\mu}$ we can define a normal projector $N^{\mu}{}_{\nu}=N^{\mu}N_{\nu}$ and a tangential projector $P^{\mu}{}_{\nu} = \delta^{\mu}_{\nu}-N^{\mu}{}_{\nu}$.

A natural question is what sort of geometry the bulk NC geometry induces on $\partial\mathcal{M}$. The answer to that question depends on whether
\be
n_{\perp} \equiv n_{\mu} N^{\mu}\,,
\ee
is zero or nonzero. If $n_{\perp}=0$, then the pullback of $g_{\mu\nu}$ is degenerate and $(n_{\alpha},g_{\alpha\beta})$ give the basic building blocks for a NC geometry on $\partial\mathcal{M}$. However, if $n_{\perp}\neq 0$, then the pullback of $g_{\mu\nu}$ is a positive tensor and so $g_{\alpha\beta}$ gives a Riemannian metric on $\partial\mathcal{M}$. 

In the main text we had $n = dt$, $g_{t\mu}=0$, and further the boundary was time-independent, so that $n_{\perp}=0$. We address the most general scenario in this Appendix. To do so we find it convenient to work with the embedding functions and the connection coefficients $\Gamma^{\mu}{}_{\nu\rho}$, rather than the frame fields and spin connection as we did in the main text.

We proceed by defining a derivative on $\partial\mathcal{M}$, which we call $\mathring{D}_{\alpha}$. $\mathring{D}_{\alpha}$ can act on tensors which have both boundary and bulk indices. For example, on a tensor $\mathfrak{U}^{\mu}_{\alpha}$ with both bulk and boundary indices it acts as
\be
\mathring{D}_{\alpha}\mathfrak{U}^{\mu}_{\beta} = \partial_{\alpha} \mathfrak{U}^{\mu}_{\beta} + \Gamma^{\mu}{}_{\nu\alpha}\mathfrak{U}^{\nu}_{\beta} - \mathring{\Gamma}^{\gamma}{}_{\beta\alpha}\mathfrak{U}^{\mu}_{\gamma}\,,
\ee
where
\begin{align}
\begin{split}
\Gamma^{\mu}{}_{\nu\alpha} &= \Gamma^{\mu}{}_{\nu\rho}f^{\rho}_{\alpha}\,,
\\
\mathring{\Gamma}^{\alpha}{}_{\beta\gamma} & = f^{\alpha}_{\mu}\partial_{\gamma}f_{\beta}^{\mu} + f^{\alpha}_{\mu} \Gamma^{\mu}{}_{\nu\gamma}f^{\nu}_{\beta}\,.
\end{split}
\end{align}
The derivative of the $f_{\alpha}^{\mu}$ defines the second fundamental form $\II^{\mu}{}_{\alpha\beta}$,
\be
\II^{\mu}{}_{\alpha\beta} \equiv \mathring{D}_{\beta}f^{\mu}_{\alpha}\,.
\ee

This derivative has several useful properties. The ones we need are
\begin{align}
\begin{split}
\mathring{D}_{\alpha}n_{\mu} &= 0\,, \quad \mathring{D}_{\alpha}g_{\mu\nu}=0\,,
\\
f_{\mu}^{\alpha} \II^{\mu}{}_{\beta\gamma} & = 0\,, \quad \mathring{D}_{\alpha}\gamma_{\beta\gamma}  = 0 \,.
\end{split}
\end{align}
In particular, this implies that $\II^{\mu}{}_{\alpha\beta}$ satisfies $\II^{\mu}{}_{\alpha\beta} = N^{\mu} k_{\alpha\beta}$ for some tensor $k_{\alpha\beta}$. From this we define the extrinsic curvature $\bar{K}_{\alpha\beta}$ via 
\be
\label{E:barK}
\II^{\mu}{}_{\alpha\beta} = \frac{N^{\mu}}{1-n_{\perp}^2}\bar{K}_{\alpha\beta}\,,
\ee
or equivalently using $N^{\mu}N^{\nu}g_{\mu\nu}=1-n_{\perp}^2$
\be
\label{E:barK2}
\bar{K}_{\alpha\beta} = N^{\mu}g_{\mu\nu}\II^{\nu}{}_{\alpha\beta}= -(1-n_{\perp}^2) f^{\mu}_{\alpha}\mathring{D}_{\beta}N_{\mu}\,.
\ee
In general, $\bar{K}_{\alpha\beta}$ has an antisymmetric part owing to the torsion. It is also useful to define an ``unnormalized'' extrinsic curvature $\mathcal{K}_{\alpha\beta}=N_{\mu}\II^{\mu}{}_{\alpha\beta}$ which is related to $\bar{K}_{\alpha\beta}$ by $\bar{K}_{\alpha\beta}=(1-n_{\perp}^2)\mathcal{K}_{\alpha\beta}$.

There are two curvatures one can build from $\mathring{D}_{\alpha}$. In terms of the connection one-forms $\bar{\Gamma}^{\mu}{}_{\nu} \equiv \text{P}[\Gamma^{\mu}{}_{\nu}]=\Gamma^{\mu}{}_{\nu\alpha}d\sigma^{\alpha}$ and $\mathring{\Gamma}^{\alpha}{}_{\beta} = \mathring{\Gamma}^{\alpha}{}_{\beta\gamma}d\sigma^{\gamma}$, they are
\begin{align}
\begin{split}
\bar{R}^{\mu}{}_{\nu} &= d\bar{\Gamma}^{\mu}{}_{\nu} + \bar{\Gamma}^{\mu}{}_{\rho}\wedge \bar{\Gamma}^{\rho}{}_{\nu}\,,
\\
\mathring{R}^{\alpha}{}_{\beta} & = d\mathring{\Gamma}^{\alpha}{}_{\beta} + \mathring{\Gamma}^{\alpha}{}_{\gamma}\wedge \mathring{\Gamma}^{\gamma}{}_{\beta}\,.
\end{split}
\end{align}
The barred curvature is nothing more than the pullback of $R^{\mu}{}_{\nu}$,
\be
\bar{R}^{\mu}{}_{\nu} = \text{P}[R^{\mu}{}_{\nu}]\,.
\ee
The $\bar{R}^{\mu}{}_{\nu}$ and $\mathring{R}^{\alpha}{}_{\beta}$ are related to each other and the extrinsic curvature by the NC analogue of the Gauss, Codazzi, and Ricci equations, which we now derive.

As at the end of the previous Appendix, we define an exterior covariant derivative $\mathring{D}$. For any vector field $\mathfrak{v}^{\mu}$ restricted to $\partial\mathcal{M}$ and vector field $\mathfrak{w}^{\alpha}$ on $\partial\mathcal{M}$ it satisfies
\be
\label{E:Dsq}
\mathring{D}^2\mathfrak{v}^{\mu} = \bar{R}^{\mu}{}_{\nu}\mathfrak{v}^{\nu}\,, \quad \mathring{D}^2 \mathfrak{w}^{\alpha}=\mathring{R}^{\alpha}{}_{\beta}\mathfrak{w}^{\beta}\,.
\ee
Decomposing $\mathfrak{v}^{\mu}$ into normal and tangential parts as
\be
\label{E:decomposev}
\mathfrak{v}^{\mu} = f^{\mu}_{\alpha} \mathfrak{v}^{\alpha} + \mathfrak{v}_{\perp} N^{\mu}\,,
\ee
its derivative has tangential and normal parts,
\be
\mathring{D}\mathfrak{v}^{\mu} = f^{\mu}_{\alpha}\left(\mathring{D}\mathfrak{v}^{\alpha} - \mathcal{K}^{\alpha}\mathfrak{v}_{\perp}\right) + N^{\mu}\left( \mathring{D}\mathfrak{v}_{\perp} + \mathcal{K}_{\alpha}\mathfrak{v}^{\alpha}\right)\,,
\ee
where we have defined $\mathcal{K}_{\alpha}=\mathcal{K}_{\alpha\beta}d\sigma^{\beta}$ and $\mathcal{K}^{\alpha}=\gamma^{\alpha\beta}\mathcal{K}_{\beta}$. Taking a second derivative gives
\begin{align}
\begin{split}
\label{E:onTheOneHand}
\mathring{D}^2 \mathfrak{v}^{\mu} = & f^{\mu}_{\alpha}\left( \mathring{R}^{\alpha}{}_{\beta} \mathfrak{v}^{\beta} - \mathcal{K}^{\alpha}\wedge \mathcal{K}_{\beta}\mathfrak{v}^{\beta}- \mathring{D}\mathcal{K}^{\alpha} v_{\perp} \right)
\\
& +N^{\mu}\left(\mathring{D} \mathcal{K}_{\alpha}\mathfrak{v}^{\alpha} - \mathcal{K}_{\alpha}\wedge \mathcal{K}^{\alpha} \mathfrak{v}_{\perp}\right)\,.
\end{split}
\end{align}
We also find, by substituting~\eqref{E:decomposev} into~\eqref{E:Dsq},
\be
\label{E:onTheOther}
\mathring{D}^2 \mathfrak{v}^{\mu} = \bar{R}^{\mu}{}_{\nu}f^{\nu}_{\alpha}\mathfrak{v}^{\alpha} + \bar{R}^{\mu}{}_{\nu}N^{\nu}\mathfrak{v}_{\perp}\,.
\ee
Comparing these expressions gives
\begin{align}
\begin{split}
\label{E:gaussCodazziRicci}
f^{\alpha}_{\mu}\bar{R}^{\mu}{}_{\nu}f^{\nu}_{\beta} &= \mathring{R}^{\alpha}{}_{\beta}-\mathcal{K}^{\alpha}\wedge \mathcal{K}_{\beta} \,,
\\
N_{\mu}\bar{R}^{\mu}{}_{\nu}f^{\nu}_{\alpha} & = \mathring{D}\mathcal{K}_{\alpha}\,,
\\
f^{\alpha}_{\mu}\bar{R}^{\mu}{}_{\nu}N^{\nu} & = - \mathring{D}\mathcal{K}^{\alpha}\,,
\\
N_{\mu}\bar{R}^{\mu}{}_{\nu}N^{\nu} & = - \mathcal{K}^{\alpha}\wedge\mathcal{K}_{\alpha}\,.
\end{split}
\end{align}
The first of these equations is analogous to the Gauss equation, the second and third to the Codazzi equation, and the last to the Ricci equation.

The relations~\eqref{E:gaussCodazziRicci} can be nicely summarized in the following way. Define the matrix-valued one-form
\be
\mathcal{M}^{\mu}{}_{\nu} \equiv N^{\mu}\mathcal{K}_{\alpha}f^{\alpha}_{\nu} - f^{\mu}_{\alpha}\mathcal{K}^{\alpha}N_{\nu}\,,
\ee
as well as a new connection
\be
\tilde{\Gamma}^{\mu}{}_{\nu} \equiv \bar{\Gamma}^{\mu}{}_{\nu}-\mathcal{M}^{\mu}{}_{\nu}\,.
\ee
The curvature of $\tilde{\Gamma}^{\mu}{}_{\nu}$, $\tilde{R}^{\mu}{}_{\nu} = d\tilde{\Gamma}^{\mu}{}_{\nu} + \tilde{\Gamma}^{\mu}{}_{\rho}\wedge \tilde{\Gamma}^{\rho}{}_{\nu}$, is
\be
\label{E:betterGCR}
\tilde{R}^{\mu}{}_{\nu} = f^{\mu}_{\alpha}f_{\nu}^{\beta} \mathring{R}^{\alpha}{}_{\beta}\,,
\ee
which is equivalent to~\eqref{E:gaussCodazziRicci} upon expressing the LHS as
\be
\tilde{R}^{\mu}{}_{\nu} = \bar{R}^{\mu}{}_{\nu}-\mathring{D}\mathcal{M}^{\mu}{}_{\nu} + \mathcal{M}^{\mu}{}_{\rho}\wedge \mathcal{M}^{\rho}{}_{\nu}\,.
\ee
We observe that there is an obvious generalization of~\eqref{E:betterGCR} for Riemannian manifolds with boundary, which we have not seen in the literature.

So much for $\bar{R}^{\mu}{}_{\nu}$. Specializing to $d=2$, we would like to express $\text{P}[\mathcal{R}]$ in terms of the boundary data. A straightforward computation using~\eqref{E:gaussCodazziRicci},
\be
n_{\mu}R^{\mu}{}_{\nu}=-D^2 n_{\nu}=0\,,
\ee
and $\varepsilon^{\alpha\beta} = N_{\mu}f^{\alpha}_{\nu}f^{\beta}_{\rho}\varepsilon^{\mu\nu\rho}$ shows that
\be
\text{P}[\mathcal{R}] = \frac{1}{2}\varepsilon^{\mu\nu\rho}n_{\mu}\bar{R}_{\nu\rho} = -d \left( \varepsilon^{\alpha\beta}n_{\alpha}\bar{K}_{\beta}\right)\,,
\ee
where $\bar{K}_{\beta}=\bar{K}_{\beta\gamma}d\sigma^{\gamma}$ and $\bar{K}_{\beta\gamma}$ is the normalized extrinsic curvature defined in~\eqref{E:barK}. 

Now define the one-form in brackets to be
\be
K \equiv \varepsilon^{\alpha\beta}n_{\alpha}\bar{K}_{\beta}\,.
\ee
Since $\mathcal{R}=d\omega$, it follows that
\begin{align}
\begin{split}
&\int_{\mathcal{M}} A \wedge d\omega + \int_{\partial\mathcal{M}} A \wedge K\,,
\\
& \int_{\mathcal{M}}\omega \wedge d\omega + \int_{\partial\mathcal{M}} A \wedge K\,,
\end{split}
\end{align}
are invariant under $U(1)$ gauge transformations and local $SO(2)$ rotations. Recall that this was the primary result of the main text, given in~(12) and~(13).

Let us now relate these results to the case discussed in the main text, with $n=dt, g_{t\mu}=0$ and a time-independent boundary. In that case $n_{\perp}=0$, the normal vector is spatial $N^i$, and $\varepsilon^{\beta\alpha}n_{\alpha}$ is the spatial tangent vector $t^{i}$, so that using~\eqref{E:barK2} we find
\be
K_{\alpha} = - t^{\mu}\mathring{D}_{\alpha}N_{\mu}= N_{\mu}\mathring{D}_{\alpha}t^{\mu}\,,
\ee
which is equivalent to~(8). Since we also have
\be
\text{P}[\mathcal{R}] = d\text{P}[\omega]\,,
\ee
it follows that
\be
\omega_{\alpha} + K_{\alpha} = d\varphi\,,
\ee
for $\varphi$ a locally defined function on $\partial\mathcal{M}$, which justifies~(9).

\bibliography{Bibliography}